\documentstyle[art12,bezier]{article}
\def\si{\quad}
\def\sii{\qquad}

\def\siiii{\qquad\qquad}
\def\b:{\begin{equation}}
\def\b:{\begin{equation}}
\def\b:{\begin{equation}}
\def\e:{\end{equation}}
\def\be:{\begin{eqnarray}}
\def\ee:{\end{eqnarray}}
\def\nn{\nonumber\\}
\def\cl{\centerline}

\def\vi{\vglue 1cm}
\def\vii{\vglue 2cm}

\def\la{\langle}
\def\ra{\rangle}

\def\ve{\vfill\eject}
\def\bmin{\begin{minipage}{15cm} }
\def\emin{\end{minipage}}
\newcommand{\lb}[1]{\label{eqn:#1}}
\newcommand{\rf}[1]{\ref{eqn:#1}}
\begin{document}
\baselineskip 18pt
\vii

\cl{{\Large {\bf The Correspondence between Discrete Surface}}}

\cl{{\Large {\bf and Difference Geometry of the KP-hierarchy\footnote{
This work is supported in part by the Grant-in-Aid for general Scientific
Research from the Ministry of Education, Science, Sports and Culture, Japan
(No.06835023), and the Fiscal Year 1996 Fund for Special Research Projects
at Tokyo Metropolitan University.}}}}

\vi
\cl{Satoru SAITO}
\vi
\cl{\it Depertment of Physics, Tokyo Metropolitan University,}

\cl{\it  Hachiohji, Tokyo 192-03 Japan }

\vglue 10cm
\begin{abstract}
The correspondence between two geometrical descriptions of the KP-hierarchy,
one by discrete surface and another by difference analogue of differential
geometry, is given.
\end{abstract}

PACS codes: 02.20.Tw \  02.40.-k

Keywords: Difference geometry, KP-hierarchy, Moyal quantisation, Hirota
equation
\ve
In the traditional physics the fundamental laws of nature were described in
terms of differential equations. Their corresponding discrete versions were
considered being approximation appropriate to compute them numerically or to
avoid divergence. If there exists an integrable discrete system, on the other
hand, the situation becomes completely different.  A single discrete integrable
equation can be associated to infinitly many continuous ones. It will be more
natural to consider the discrete integrable equation being more fundamental,
rather than special deformations of continuous version.

Among other discrete integrable systems the Hirota bilinear difference
equation, which we abbreviate as HBDE\cite{Hirota}, or 2D Toda equation, plays
an important role in theoretical physics. This single equation describes the
totality of soliton equations in the KP hierarchy. The connections of this
equation were shown to algebraic curves\cite{Sato}, to the string correlation
function in particle physics\cite{S}, to the 2D Ising model\cite{McCoy}, to the
Yang-Baxter equation of statistical lattice
models\cite{Kuniba},\cite{Krichever}, to the Painlev\'e system\cite{Ramani}, to
the completely integrable cellular automatons\cite{Tokihiro}.

Very recently HBDE has been given a geometrical interpretation by means of
discrete geometry\cite{Doliwa},\cite{Bobenko} of two dimensional surface. There
the HBDE turns out to be a recurrence formula satisfied by projective
invariants of the Laplace sequence of the discrete analogue of the conjugate
nets. On the other hand we developed a formulation of difference analogue of
differential geometry\cite{KS}. The purpose of our work was to describe the
Moyal quantisation as a discretisation of phase space and to provide an overall
view to integrable systems. To this end we defined difference analogue of
vector fields, differential forms, Lie derivatives and so on. In
ref.\cite{Kemmoku} Kemmoku showed that this formalism naturally induces the
infinite symmetry which characterises the KP-system. These two works, one of
geometry of discrete surface and the other of difference analogue of
differential geometry, were done quite independently. In fact they have nothing
in common apart from the fact that both describe discrete integrable systems.

Our question we discuss in this article is if they are related. The answer is
yes. In fact they are the same.

Let us begin with writing HBDE:
\be:
&&\alpha f_n(l+1,m)f_n(l,m+1)+\beta f_n(l,m)f_n(l+1,m+1)\nn
&&\siiii +\gamma f_{n+1}(l+1,m)f_{n-1}(l,m+1)=0.
\lb{HBDE}
\ee:
where $l,m,n\in {\mbox{\boldmath$Z$}}$ are discrete variables, $\alpha, \beta,
\gamma \in {\mbox{\boldmath$C$}}$ are parameters subject to the constraint
$\alpha+\beta+\gamma=0$.

The statement in ref.\cite{Doliwa} is as follows. We consider in the 3D
Euclidian space a quadrilateral $\la Y(l,m),Y(l+1,m),Y(l,m+1),Y(l+1,m+1)\ra$,
called elementary quadrilateral. If they satisfy the discrete Laplace equation
\b:
Y(l+1,m+1)-Y(l,m)=A(l,m)(Y(l+1,m)-Y(l,m))+B(l,m)(Y(l,m+1)-Y(l,m)),
\lb{Laplace map}
\e:
they form a planar quadrilateral. A collection of such planar quadrilaterals
$Y$ is called a discrete conjugate net. The Laplace transform of the net
$Y\equiv Y_0$ defines the other two conjugate nets $Y_1$ and $Y_{-1}$ by
connecting the intersection points of the directions of the opposite edges of
the four elementary quadrilaterals which intersect at $Y_0(l,m)$. The same
process can be repeated to the images of the map, hence we obtain a sequence of
conjugate nets, $Y_n,\ n=0,\pm 1,\pm 2,\cdots$. At every step of the map there
exists an invariant under the projective transformations, called a cross ratio
$f_n(l,m)$, which is formed by the four points $\{Y_n(l,m-1),Y_n(l,m),
Y_{n+1}(l,m), Y_{n-1}(l-1,m)\}$ of the $n$th map. Since the cross ratio is
determined by the coefficients $A_n$ and $B_n$ of the Laplace map of the form
$(\rf{Laplace map})$, the sequence of the maps define a recurrence formula
among the cross ratios. This formula was shown in ref.\cite{Doliwa} being
equivalent to the HBDE after some normalisation. We notice that this sequence
of the map is exactly the same one found in ref.\cite{SS3} in connection with
the linearisation of HBDE.

The difference analogue of differential geometry was discussed in ref.\cite{KS}
as follows. Let \mbox{\boldmath$x$} and \mbox{\boldmath$a$} be vectors in
${\mbox{\boldmath$R$}}^n$ and define a difference vector field associated with
a ${\mbox{\boldmath$C$}}^\infty$ function
$v({\mbox{\boldmath$x$}},{\mbox{\boldmath$a$}})$ by
\b:
{\mbox{\boldmath$X$}}^D:=\int d{\mbox{\boldmath$a$}}\
v({\mbox{\boldmath$x$}},{\mbox{\boldmath$a$}})\nabla_{\mbox{\boldmath$a$}},
\e:
where the difference operator is given by
\b:
\nabla_{\mbox{\boldmath$a$}}:={1\over\lambda}
\sinh(\lambda\sum_ja_j\partial_{x^j}).
\e:
We also define the operator $\Delta^{\mbox{\boldmath$a$}}$ dual of
$\nabla_{\mbox{\boldmath$a$}}$ under the following inner product
\b:
\la \Delta^{\mbox{\boldmath$a$}}|
\nabla_{\mbox{\boldmath$b$}}\ra=\delta({\mbox{\boldmath$a$}}-
{\mbox{\boldmath$b$}}).
\e:
Then a difference one form is defined by (see ref.\cite{KS} for more detail)
\b:
\Omega^D:=\int d{\mbox{\boldmath$a$}}\
w({\mbox{\boldmath$x$}},{\mbox{\boldmath$a$}})\Delta^{\mbox{\boldmath$a$}},
\e:
which gives rise to an invariant upon taking the inner product with a vector
field:
\b:
\la \Omega^D|{\mbox{\boldmath$X$}}^D\ra=\int d{\mbox{\boldmath$x$}}\int
d{\mbox{\boldmath$a$}}\
w({\mbox{\boldmath$x$}},{\mbox{\boldmath$a$}})v({\mbox{\boldmath$x$}},
{\mbox{\boldmath$a$}}).
\e:

Extending the space to the $2n$ dimensional phase space,
$({\mbox{\boldmath$x$}},{\mbox{\boldmath$p$}}) \in
{\mbox{\boldmath$R$}}^n\times {\mbox{\boldmath$R$}}^n$, a difference analogue
of Hamiltonian vector field is defined as
\b:
{\mbox{\boldmath$X$}}_A^D:=\int
d{\mbox{\boldmath$a$}}_1d{\mbox{\boldmath$a$}}_2\
v_A({\mbox{\boldmath$x$}},{\mbox{\boldmath$p$}};{\mbox{\boldmath$a_1$}},,
{\mbox{\boldmath$a$}}_2)\nabla_{{\mbox{\boldmath$a$}}_1,
{\mbox{\boldmath$a$}}_2},
\lb{Hamiltonian vector field}
\e:
where
\b:
v_A({\mbox{\boldmath$x$}},{\mbox{\boldmath$p$}};{\mbox{\boldmath$a_1$}},
{\mbox{\boldmath$a$}}_2)=\left({\lambda\over 2\pi}\right)^{2n}
\int d{\mbox{\boldmath$b$}}_1d{\mbox{\boldmath$b$}}_2\
 e^{-i({\mbox{\boldmath$a_1$}}{\mbox{\boldmath$b$}}_2-
{\mbox{\boldmath$a$}}_2{\mbox{\boldmath$b$}}_1)}A({\mbox{\boldmath$x$}
}+\lambda {\mbox{\boldmath$b$}}_1,{\mbox{\boldmath$p$}
}+\lambda{\mbox{\boldmath$b$}}_2).
\lb{symplectic}
\e:
The operation of the Hamiltonian vector field to a function on the phase space
yields the Moyal bracket\cite{Moyal} :
\be:
{\mbox{\boldmath$X$}}_A^DB({\mbox{\boldmath$x$}},{\mbox{\boldmath$p$}})
&=&\left.{1\over\lambda}\sin\left\{\lambda\left(\partial_{
{\mbox{\boldmath$x$}}_A}\partial_{{\mbox{\boldmath$p$}}_B}-
\partial_{{\mbox{\boldmath$x$}}_B}\partial_{{\mbox{\boldmath$p$}}_A
}\right)\right\}A({\mbox{\boldmath$x$}}_A,{\mbox{\boldmath$p$}}_A
)B({\mbox{\boldmath$x$}}_B,{\mbox{\boldmath$p$}}_B)
\right|_{{\mbox{\boldmath$x$}},{\mbox{\boldmath$p$}}}\nn
&=:&i\{A,B\}_M.
\ee:
It is apparent from this expression that in the continuous limit of the phase
space $\lambda\rightarrow 0$ we recover the classical Poisson bracket.

The above definition of the Hamiltonian vector fields enables us to extend the
concept of the Lie derivative in the differential geometry to the discrete
phase space. In fact we can show directly the following operator relation
between two vector fields:
\b:
\left[{\mbox{\boldmath$X$}}_A^D,{\mbox{\boldmath$X$}}_B^D\right]=,
{\mbox{\boldmath$X$}}^D_{-i\{A,B\}_M}.
\lb{Moyal algebra}
\e:
This exhibits a large symmetry generated by the Hamiltonian vector fields,
which should be common in the physical systems formulated in our prescription.
This algebra itself was discussed\cite{Fairlie}\cite{Floratos} in various
contexts including some geometrical arguments.

As we consider a physical phase space the pairing of a vector field with a
difference one form is expected to mean an expectation value of a physical
quantity. It is accomplished by adopting the Wigner distribution
function\cite{Wigner} $F({\mbox{\boldmath$x$}},{\mbox{\boldmath$p$}})$ to
specify the physical state and by defining the difference one form by
\b:
\Omega^D_F:=\int d{\mbox{\boldmath$a$}}_1d{\mbox{\boldmath$a$}}_2\ \int
d{\mbox{\boldmath$b$}}_1d{\mbox{\boldmath$b$}}_2\
e^{i({\mbox{\boldmath$a_1$}}{\mbox{\boldmath$b$}}_2-
{\mbox{\boldmath$a$}}_2{\mbox{\boldmath$b$}}_1)}
F({\mbox{\boldmath$x$}}+\lambda {\mbox{\boldmath$b$}}_1
,{\mbox{\boldmath$p$}}+\lambda{\mbox{\boldmath$b$}}_2)
\Delta^{{\mbox{\boldmath$a$}}_1,{\mbox{\boldmath$a$}}_2}.
\e:
The expectation value of a physical quantity, say
$A({\mbox{\boldmath$x$}},{\mbox{\boldmath$p$}})$, is obtained by calculating
the inner product $\la \Omega^D_F|{\mbox{\boldmath$X$}}_A^D\ra$.

Two types of prescriptions of discrete analogue of differential geometry are
explained in above briefly. We notice that in the geometrical approach a real
discretisation of two dimensional surface is investigated while in the other
approach a formal discretisation of phase space with arbitrary dimensionality
is the subject of the study. The space which is discretised is different.
Moreover the meaning of discretisation of the phase space do not correspond to
real discretisation of \mbox{\boldmath$x$} and \mbox{\boldmath$p$} separately,
since only their product must have a finite minimum $\lambda$. In fact
uncertainty relations can be proved for the product of their square mean
values\cite{Moyal}. There seems little hope to find direct correspondence
between these two discretisations.

What we like to discuss in the following is that, contrary to this naive
observation, they are simply different view of the same object. For this
purpose let us recall the two different view of the KP-hierarchy. One is given
by $(\rf{HBDE})$ in which three discrete variables $l,m,n$ are used. The other
representation is possible by using the KP-flow parameters $\{t_n|n\in
{\mbox{\boldmath$N$}}\}$ and splitting $(\rf{HBDE})$ into many soliton
equations through reduction of variables. The variables are related each other
by the Miwa transformation\cite{Miwa}:
\b:
t_n={1\over n}\sum_{j}k_jz_j^n,\sii n\in {\mbox{\boldmath$N$}}
\lb{Miwa}
\e:
where $z_j$'s are arbitrary parameters and
\b:
l=p+q+r-{3\over 2},\si m=-p-{1\over 2},\si n=p+q-1
\e:
with $(p,q,r)$ being arbitrary three out of $k_j$'s in $(\rf{Miwa})$. We know
that the geometry of the discrete surface was described by means of the shift
operators $e^{\partial_l}, e^{\partial_m}$, whereas the third shift operator
$e^{\partial_n}$ generates the sequence of Laplace transforms. The question is
if they correspond to some difference vector fields in the form of
$(\rf{Hamiltonian vector field})$ when they are translated into the language of
the other variables $t_n$'s.

After the change of variables according to $(\rf{Miwa})$ we have
\b:
e^{\partial_{k_j}}=\exp\left[\sum_{n=1}^\infty{1\over
n}z_j^n{\partial\over\partial t_n}\right].
\e:
The right hand side can be interpreted as a simultaneous shifts to all
directions. The amount of the shift along the $t_n$ direction is ${1\over
n}z_j^n$. The antisymmetric combination of $e^{\partial_{k_j}}$ forms a vector
field with a constant weight function. Since $l,m,n$ are linear combinations of
$k_j$'s, the shifts to these directions are produced by the shifts to the $t_n$
directions by ${1\over n}(z_p^n+z_q^n+z_r^n),\ -{1\over n}z_p^n,\ {1\over
n}(z_p^n+z_q^n)$ respectively.

{}From this consideration it is convenient to introduce a shift vector field:
\b:
{\mbox{\boldmath$X$}}^S_A:=\int d{\mbox{\boldmath$a$}}\
A({\mbox{\boldmath$t$}},{\mbox{\boldmath$a$}}
)e^{\lambda\sum_na_n\partial_{t_n}}.
\lb{shift operator}
\e:
We can convince ourselves that the shift vector fields form an algebra among
themselves:
\b:
[{\mbox{\boldmath$X$}}^S_A,\
{\mbox{\boldmath$X$}}^S_B]={\mbox{\boldmath$X$}}^S_{-\{A,B\}_S},
\lb{shift algebra}
\e:
where
\be:
\{A,B\}_S&:=&\int d{\mbox{\boldmath$b$}}\
\left(\exp\left[\lambda\left({{\mbox{\boldmath$a$}}\over
2}-{\mbox{\boldmath$b$}}\right)\partial_{{\mbox{\boldmath$t$}}_A}\right
]-\exp\left[\lambda\left({{\mbox{\boldmath$a$}}\over 2}+{\mbox{\boldmath$b$}
}\right)\partial_{{\mbox{\boldmath$t$}}_B}\right]\right)\nn
&&\siiii\times\left.A\left({\mbox{\boldmath$t$}}_A,{{\mbox{\boldmath$a$}}\over
2}+{\mbox{\boldmath$b$}}\right)B\left({\mbox{\boldmath$t$}}_B,
{{\mbox{\boldmath$a$}}\over 2}-{\mbox{\boldmath$b$}}\right)
\right|_{{\mbox{\boldmath$t$}}_A={\mbox{\boldmath$t$}}_B=
{\mbox{\boldmath$t$}}}.
\ee:

We also introduce a gauge covariant shift operator, which is defined by
\be:
{\mbox{\boldmath$X$}}^S_{k_j}[U]&:=&U({\mbox{\boldmath$k$}})
e^{\partial_{k_j}}U^{-1}({\mbox{\boldmath$k$}})\nn
&=:&u({\mbox{\boldmath$t$}})e^{\sum_n{1\over n}z_j^n\partial_{t_n}}.
\ee:
Here $U({\mbox{\boldmath$k$}})$ is a function of
${\mbox{\boldmath$k$}}=\{k_1,k_2,\cdots\}$ and we defined
$u({\mbox{\boldmath$t$}})$ by
\b:
u({\mbox{\boldmath$t$}})=U({\mbox{\boldmath$k$}})U^{-1}
({\mbox{\boldmath$k$}}'),\sii {\mbox{\boldmath$k$}}'
:=\{k_1,k_2,\cdots,k_j+1,k_{j+1},\cdots\}.
\e:
The correspondence of ${\mbox{\boldmath$X$}}^S_{k_j}[U]$ to
${\mbox{\boldmath$X$}}^S_A$ of $(\rf{shift operator})$ is achieved by choosing
\b:
A({\mbox{\boldmath$t$}},{\mbox{\boldmath$a$}})=u({\mbox{\boldmath$t$}})
\prod_n\delta\left(a_n-{1\over\lambda n}z_j^n\right).
\e:
These new operators enable us to translate various terms in the KP-hierarchy in
the language of $t_n$'s into the language of $k_j$'s. We will give a list of
such translation in the following.

\begin{enumerate}
\item
Vertex operator I:

First we consider the case
\b:
U({\mbox{\boldmath$k$}})=\prod_{l\ne j}\left(1-q{z_l\over z_j}\right)^{k_lk_j}
\lb{U_0}
\e:
and let
\b:
t_n={1\over n}\sum_jk_jq^nz_j^n,\sii |q|<1,\si n=1,2,\cdots.
\e:
Then we find
\be:
{\mbox{\boldmath$X$}}^S_{k_j}[U]&=&\prod_{l\ne j}\left(1-q{z_l\over
z_j}\right)^{-k_l}e^{\partial_{k_j}}\nn
&=&(1-q)^{k_j}\exp\left[\sum_{n=1}^\infty
t_nz_j^{-n}\right]\exp\left[\sum_{n=1}^\infty{1\over
n}q^nz_j^n{\partial\over\partial t_n}\right].
\ee:
This last expression, after devided by the factor $(1-q)^{k_j}$ and taking the
$q\rightarrow 1$ limit, is the one called the vertex operator in the soliton
theory.
\item
Vertex operator II:

We can generalise the vertex operator by making double the space of $t_n$'s.
For this we define
\b:
U({\mbox{\boldmath$k$}}):=\prod_{l\ne j}\left(1-{\bar z_l\over
z_j}\right)^{k_lk_j}\left(1-{\bar z_j\over z_l}\right)^{-k_lk_j}
\e:
and
\be:
t_n&:=&{1\over n}\sum_jk_jz^n_j,\si n=1,2,\cdots,\sii t_0:=\sum_jk_j\ln z_j\nn
\bar t_n&:=&{1\over n}\sum_jk_j\bar z^{-n}_j,\si n=1,2,\cdots,\sii \bar
t_0:=-\sum_jk_j\ln \bar z_j,
\ee:
where we added the 0-th components $t_0$ and $\bar t_0$ as well. In terms of
$t_n$'s
\be:
{\mbox{\boldmath$X$}}^S_{k_j}[U]
&=&\exp\left[-\sum_{n=0}^\infty(t_n\bar z^{-n}_j-\bar t_nz^n_j)\right]\nn
&&\times\exp\left[\ln z_j\partial_{t_0}-\ln \bar z_j\partial_{\bar
t_0}+\sum_{n=1}^\infty{1\over n}(z^n_j\partial_{t_n}+\bar
z^{-n}_j\partial_{\bar t_n})\right]\nn
&=:&e^{iX_+(z_j,\bar z_j)}e^{iX_-(z_j,\bar z_j)}.
\ee:
The last expression is given to remind the relation of the operator to the
string coordinates $X_\pm$  \cite{S}. Note that this has a symplectic sturcture
such that $(t_n, \bar t_n)$ form a symplectic pair of the infinite dimensional
phase space as it is implied in $(\rf{symplectic})$. Therefore it satisfies the
Moyal algebra $(\rf{Moyal algebra})$, rather than $(\rf{shift algebra})$.
\item
B\"acklund transformation:

The B\"acklund transformation is generated by the operators of the form:
\b:
B(z_j,z_l)={\mbox{\boldmath$X$}}^S_{k_j}[U]{\mbox{\boldmath$X$}}^S_{-k_l}[U].
\e:
They form an algebra which follows to $(\rf{Moyal algebra})$.
\item
Linearisation of HBDE:

HBDE can be derived as an integrability condition of the following linear
problem\cite{SS1}:
\be:
{\mbox{\boldmath$X$}}^S_l[U]|\Phi\ra&=&c_+{\mbox{\boldmath$X$}}^S_{l+n}
[V]|\Phi\ra\nn
{\mbox{\boldmath$X$}}^S_m[V]|\Phi\ra&=&c_-{\mbox{\boldmath$X$}}^S_{m-n
}[U]|\Phi\ra
\ee:
It was shown in ref.\cite{SS1} that under the gauge condition
$V(l,m,n)=U(l,m,n-1)$, the gauge potential $U(l,m,n)$ satisfies HBDE
$(\rf{HBDE})$ with $\gamma=-c_+c_-\alpha$. The gauge potential $U$ in the
covariant shift operator has an obvious geometrical meaning as a connection.
HBDE is nothing but a geometrical constraint imposed on the connection. We have
also shown that this set of linear equations reduce to the Lax type of equation
for the Toda lattice in an appropriate continuous limit\cite{Nakajima Saito}.
It is apparent that writing the equations by using $t_n$'s we will get
infinitely many inverse problems depending on the way of reduction of the
variables.

\item
Baker-Akhiezer function:

The Baker-Akhiezer function of the KP-hierarchy was given explicitly in terms
of the solution $\tau$ of HBDE. It can be interpreted within our formulation as
a wave function in the form:

\b:
|\Psi(z_j)\ra={\mbox{\boldmath$X$}}^S_{k_j}[\tau^{-1}U_0]|0\ra
\e:
where $U_0$ is the one of $(\rf{U_0})$.
\end{enumerate}

We have shown various objects in the KP theory expressed differently by using
two different coordinates, one of discrete surface $k_j$'s and the other the
soliton coordinates $t_n$'s. In all examples they are translated from one to
the other by the operator ${\mbox{\boldmath$X$}}^S_{k_j}[U]$. This operator
itself possesses two faces, one is a gauge covariant shift operator in the
former coordinates. At the same time it represents a difference analogue of
vector field in the latter. If we let $U$ be an arbitrary function of $t_n$'s
or $k_j$'s, it generates a large symmetry algebra as given by $(\rf{Moyal
algebra})$. From the view point of geometry this algebra should describe some
topologial nature of the discrete surface as $k_j$'s are restricted to $p,q,r$.
An extension to higher dimensional hyper surface must be straight forward. We
know that these two sets of coordinates, $t_n$'s and $k_j$'s, are linearly
dependent each other by the Miwa transformation $(\rf{Miwa})$. Nevertheless it
is still not known what is the meaning of $k_j$'s in the space of soliton
coordinates and what is the role played by $t_n$'s on the discrete surface.

Besides two approaches discussed in this paper there exist some other discrete
approaches to the KP hierarchy or its
generalisations\cite{Nijhoff},\cite{Bogdanov}. There have been also many works
which explore geometrical nature of Moyal quantisation
\cite{Strachan}\cite{Bayen}\cite{Takasaki}\cite{Osborn}. An interpretation of
the analytical approach to rather general integrable systems \cite{Bogdanov} by
means of geometry of higher dimensional surfaces was pointed out recently in
ref.\cite{Doliwa Santini}. It is desirable to clarify correlation among these
different approaches and get insight underlying all discrete integrable
systems.

In the forthcoming paper we will discuss the symmetry generated by the
covariant shift operators and also their connection to the Yang-Baxter relation
from the view we discussed recently\cite{S2}.

The author would like to thank Prof. Y.Ohnita for pointing out
ref.\cite{Bobenko}, to Prof. T.Tokihiro ref.\cite{Doliwa Santini}, to Dr.
K.Yoshida ref.\cite{Doliwa}. He would also like to thank Drs. R.Kemmoku and
S.Matsutani for discussions and comments.

\baselineskip 15pt
{\small

\end{document}